# Differentiable Set Operations for Algebraic Expressions

Jasdeep S. Grover
K.J. Somaiya College of Engineering, Vidyavihar, Mumbai
jasdeepsingh.g@somaiya.edu

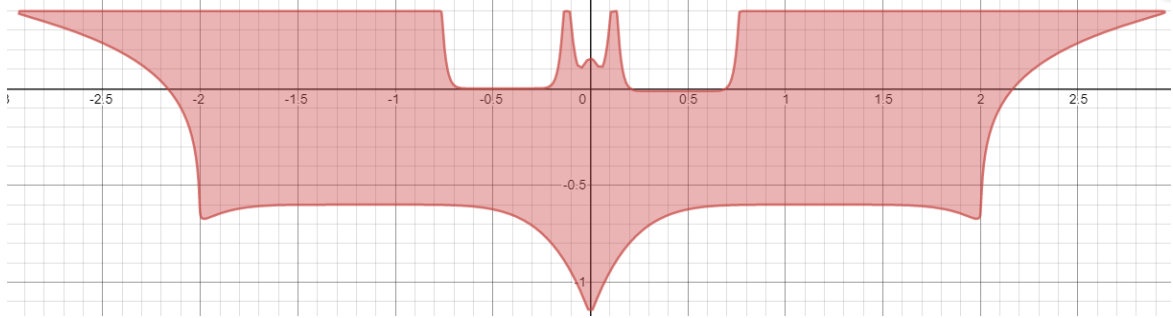

Fig 1: The Batman's Expression as shown in Eq. 1

$$e^{\{(50(-(y+1.5)))\}} + e^{\left\{\left(50\left(\left(3(y-0.1)-258.18((1.9x+0.1)(1.9x-0.1))^{\{1.6\}}\right)\right)\right)\right\}} + e^{\left\{\left(50\left(-\left((3(x-0.45))^{\{14\}}-y\right)\right)\right)\right\}}$$
$$+ e^{\left\{\left(-50\left((3(x+0.45))^{\{14\}}-y\right)\right)\right\}} + e^{\{(50(y-0.4))\}} + e^{\left\{\left(-50\left((0.5(y+1.6))^{\{8\}}+(-x+2)\right)\right)\right\}}$$
$$+ e^{\left\{\left(-50\left((0.5(y+1.6))^{\{8\}}+(x+2)\right)\right)\right\}} + e^{\left\{\left(-50\left((0.5(x+1.16)^{\{2.8\}})^{\{2\}}+(y+0.6)\right)\right)\right\}}$$
$$+ e^{\left\{\left(-50\left((0.5(x-1.16)^{\{2.8\}})^{\{2\}}+(y+0.6)\right)\right)\right\}} \leq 1 \qquad \dots Eq.\,1$$

## Abstract:

Basic principles of set theory have been applied in the context of probability and binary computation. Applying the same principles on inequalities is less common but can be extremely beneficial in a variety of fields. This paper formulates a novel approach to directly apply set operations on inequalities to produce resultant inequalities with differentiable boundaries. The suggested approach uses inequalities of the form $E_i: f_i(x_1, x_2, .., x_n) \leq 0$ and an expression of set operations in terms of $E_i$ like, $(E_1\,and\,E_2)\,or\,E_3$, or can be in any standard form like the Conjunctive Normal Form (CNF) to produce an inequality $F(x_1, x_2, .., x_n) \leq 1$ which represents the resulting bounded region from the expressions and has a differentiable boundary. To ensure differentiability of the solution, a trade-off between representation accuracy and curvature at borders (especially corners) is made. A set of parameters is introduced which can be fine-tuned to improve the accuracy of this approach. The various applications of the suggested approach have also been discussed which range from computer graphics to modern machine learning systems to fascinating demonstrations for educational purposes (current use). A python script to parse such expressions is also provided.

AMS Subject Classification, 2010: 03E20, 00A06, 00A66

## 1. Introduction:

Set operations; union, intersection and negations have been applied on sets of numbers or conceptual elements in most cases. The same have also been translated into binary operations and probability theory to obtain equivalent results in those fields. Unfortunately lesser attempts have been made to extend these principle ideas of set theory to inequalities and graph plots. This paper thus develops operations which use bounded regions represented by inequalities as sets and produces equivalent inequalities for the result of the set operation specified. Extended expressions of these operations can also be evaluated on the above inequalities. For simplifying the analysis this paper currently utilizes only 2 dimensional data represented by (x, y).

The problem can be well stipulated as follows; given a set of inequalities, find an inequality which represents the result of the provided set of operations. For example consider Eq. 2 and Eq. 3:

$$A(x, y): x^2 + y^2 - 4 \leq 0 \quad \ldots Eq. 2$$
$$B(x, y): (x - 2.5)^2 + y^2 - 4 \leq 0 \quad \ldots Eq. 3$$

Compute an approximate solution which can be enhanced incrementally for $A(x, y) \cup B(x, y)$:
Result of this paper's suggested approach is in Eq. 4 and can be seen in Fig. 2 and Fig. 3.

$$(A \cup B)(x, y): \left(e^{\{(-a(x^{\{2\}}+y^{\{2\}}-4))\}} + e^{\{(-b((x-2.5)^{\{2\}}+y^{\{2\}}-4))\}}\right)^{\{-1\}} \leq 1 \quad \ldots Eq. 4$$

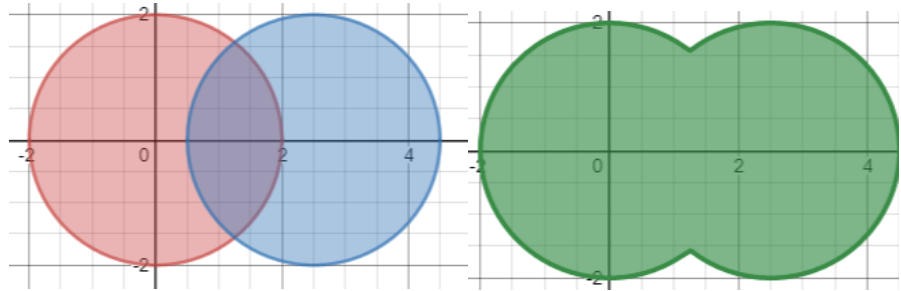

*Fig 2(left): Eq. 2(red left circle) and Eq. 3(blue right circle)*
*Fig 3(right): their union Eq. 4*
*The parameter 'a' and 'b' in Eq. 4 are positive real numbers and can be increased to increase the accuracy of the results*

The solution expected must hold the following properties:
1. Differentiability at points where all input expressions are differentiable
2. Should allow incremental enhancement to improve accuracy as differentiability is unattainable for most cases for exact representation of the resultant of set operations.

Some casual work has been done in this direction earlier. This work can be categorized into: Operations on pixels or defined units, operations on boundary points, and operations on set of equations.

I.   Operations on units generally involve identifying points in individual regions and explicitly finding results of set operations. This can be used in association with the method suggested in [1].
II.  Operations on boundaries generally tries to build a representation of the boundary of the bounded figures using various representations of the boundary like epicycles and the Fourier transform of the bounding points as discussed in [4]. The set operations can then be performed and the resultant boundary is obtained.
III. Operations on equations attempt to merge equations by developing equation segments of the form $f(x, y) = 0$. This results into a bounded region described as in Eq. 5.

$$F(f_1(x, y), f_2(x, y), \ldots f_n(x, y)) = 0 \quad \ldots Eq. 5$$

Our suggested method also falls into this category with slight modifications.

This paper then shows the various applications and implications of this result. The wide range of applications includes graphics, solutions to various challenges in regression problems and many other possibilities.

*Note that all expressions provided in this manuscript will be compatible with the Desmos platform for visualisation.*

## 2. Recent Work

This paper starts exploring with the first category of approaches using units. The Everything Formula (Tupper's Self-Referential Formula) [1] [2] as shown in Fig 5 was created to represent the set of all possible combinations of pixels hence could represent any general pattern which could be represented in the given resolution. This includes the equation itself. The given formula is again non-differentiable and cannot be used directly for evaluating set operations. Union or intersections cannot be found directly; equivalent pattern of region is always needed for computing the height of the required section which may bare the resulting pattern. The expression discussed has a unique property of plotting itself unlike the method proposed in this paper.

$$\frac{1}{2} < \left\lfloor \mathrm{mod}\left(\left\lfloor \frac{y}{17} \right\rfloor 2^{-17\lfloor x \rfloor - \mathrm{mod}(\lfloor y \rfloor, 17)}, 2\right) \right\rfloor$$

*Fig 5: Tupper's Self-referential formula and its graph [1] [2]*

Another proposed method involves representing the bounded region using epicycles and Fourier transform [3] [4] [5] as shown in Fig. 6. This can be used to identify the boundary's individual set of points, then the boundary points of the resultant can be found and a final boundary can be regenerated using methods like the Fourier Transform. This approach has been used very extensively in Image processing and in applications like character recognition. Complex methods may exist to compute the resulting boundary points but they are beyond the scope of this research. This approach does allow incremental improvement in the result but does not allow a generalized method for merging any type of functions as this paper suggested. It is specifically created for handling equalities and not inequalities.

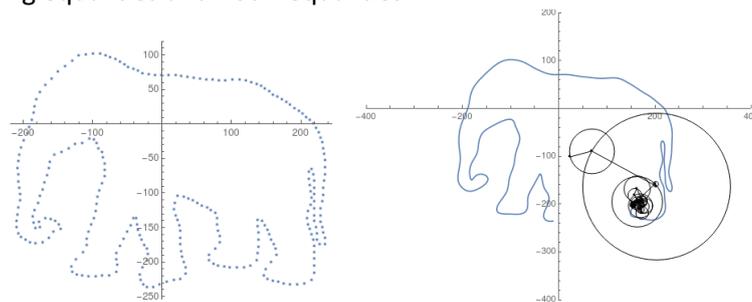

*Fig 6: Point representation of boundary and epicycles to generated the border [3] [4] [5]*

The third method actually involves combining equations in a very similar way as discussed in our approach but wasn't very well formalised. It was used in a lecture to demonstrate the batman equation [11] [6] [7] as shown in Fig. 7 and the union operation was performed using the product of functions evaluating to 0. Unfortunately details regarding the lecture were not found. The approach did not include intersection or direct use of set operations as whole. This caused the method to be somewhat restricted by the type of operations it allowed (translation of curves, rotations, scaling, clipping and merging). Again differentiability was also not assured as absolute values were used multiple times.

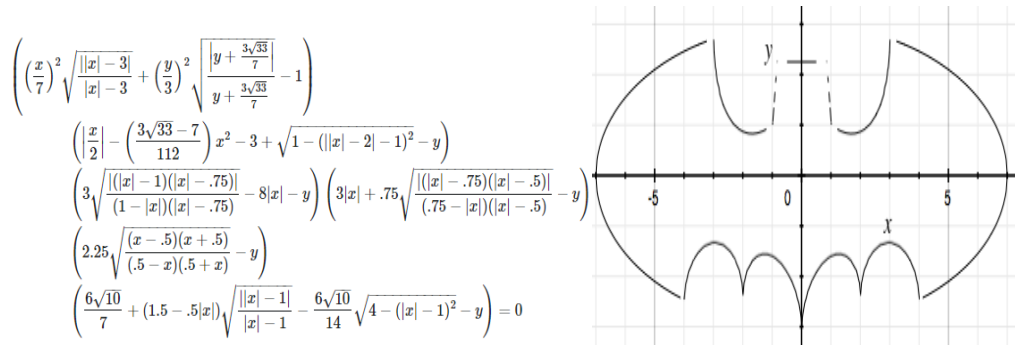

Fig 7: The Batman Equation [11] [6] [7]

Another lecture demonstration actually used a very similar approach but did not extend the ideas [8]. The idea of intersection with the use of higher even powers was also developed. Simple parallelograms were generated with the use of principles very similar to what we define here. The fact that even powers are being used directly in this approach restricts its application to certain symmetric figures and makes it computationally much more difficult to deal with. The multiplicative combination idea proposed is useful for performing XNOR but no other operations as shown in Fig. 8 and Eq. 6. Same principles were also developed independently in this paper while developing these ideas.

$$\left((x^{\{2\}} + y^{\{2\}} - 4) - 1\right)\left(((x - 2.5)^{\{2\}} + y^{\{2\}} - 4) - 1\right) \leq 0 \quad …Eq.6$$

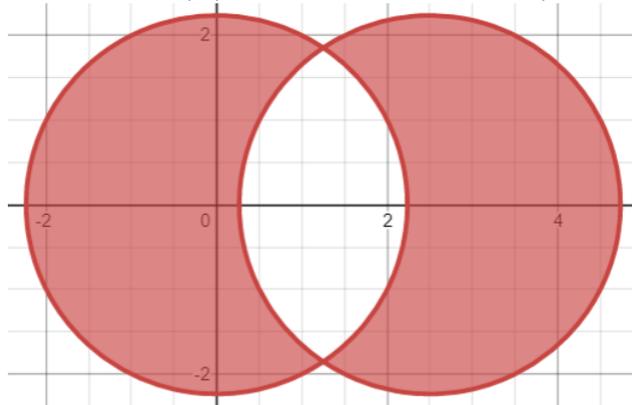

$$Fig\ 8:\ XNOR\ operation\ using\ multiplication\ in\ Eq.6$$

## 3. Proposed Method

1. ### Defining the input expressions:

    The input expressions for this method must be specified in the form of $f(x, y) \leq 0$ which is used as the standard input for the process of computing the operations on the regions. Unlike high even powers used in [8], this paper prefers using exponents to represent equations. This avoids the problems pertaining to negative solutions which interfere in the results. Hence:

    $$f(x, y) \leq 0 \ converts\ to\ e^{af(x,y)} \leq 1 \quad …Eq.7$$

    where 'a' is a parameter that can be used for determining the sharpness and accuracy of the resulting expression. Note that different values of these parameters can be associated with different components of the input. Each parameter is observed to significantly affect the region demarcated by the associated component. This representation is also compatible with the regions represented by Eq. 8 which is similar to the method in [8].

    $$f(x, y)^{2a} \leq 1\ where\ a \in Z^+ \quad …Eq.8$$

Although Eq. 8 introduces negative solutions, this method becomes more and more accurate as the value of 'a' is increased, hence can be used in conjunction with the proposed methods if needed. If $f(x, y) \leq 0$ in Eq. 8 then we can still satisfy the inequality generating solutions for negative $f(x, y)$ as well. If $f(x, y) > 1$ then the Eq. 8 should never hold true.

2. Defining Negation (Negate Exponent):

   Negation refers to the complement operation in set theory thus the region not in the inequality must be represented. We thus used the formulation in Eq. 9.
   $$e^{af(x,y)} \leq 1 \text{ being negated to } e^{-af(x,y)} \leq 1 \quad …Eq. 9$$
   Hence we take the inverse of the expression as $1/f(x, y)$. Note that boundary elements are always considered for representing the borders so equality constraint is always included. Changing the sign and other means are not considered as they might break the compatibility of the negation operation with union and intersection operations.

3. Defining Intersection (Sum):

   A very important observation which entails from [8] and observations in this paper's research is that an inequality of the form of Eq. 10 results into intersection of the equations $f_i(x, y) \leq 0$ because, if $f_i(x, y) > 0$ for any of the terms then the result will be greater than 1.
   $$e^{a_1 f_1(x,y)} + e^{a_2 f_2(x,y)} + \cdots + e^{a_n f_n(x,y)} \leq 1 \quad …Eq. 10$$
   If $f_i(x, y) < 0$ for any term then this term results into fractions which can be made to tend to 0 as $a_i$ tends to infinite. Thus increasing the parameter value generates the incremental improvement property. In most cases $a_i$ is a real number thus $e^{a_n f_n(x,y)}$ will be a small fraction which can all add up to give a number greater than 1 thus creating a source for error in the representation. Hence this method will always approximate the region represented by intersection of the individual regions but will be exact representation if parameters are infinite.

4. Defining Union(Harmonic Mean):

   Union can be defined using the De-Morgan's Law: $A \cup B \cup C \ldots = \overline{(\bar{A} \cap \bar{B} \cap \bar{C} \ldots)}$ hence producing the harmonic mean shown in Eq. 11.
   $$(e^{-a_1 f_1(x,y)} + e^{-a_2 f_2(x,y)} + \cdots + e^{-a_n f_n(x,y)})^{-1} \leq 1 \quad …Eq. 11$$

## 4. Properties of the Suggested Method

Some important derived properties of such a formulation will have significant implications as well. The above operations have the same properties as set theoretic operations provable in the case when the parameters tend to infinite. The associative and commutative properties are independent of the parameter but the distributive properties hold true only in the limiting case. This still is not a significant challenge as all expressions obtained have very small error when large integers like 50 are used as the values of the parameters, as detailed in Fig. 9. This still gives an amazing result where multiple inequalities (different before and after distribution operations) converge at the same inequality in the limiting case. We can also use the property to get multiple identical inequalities representing nearly the same region when large values of the parameters are used.

Another very significant observation which will allow us to build any shape irrespective of its complexity is that the result of the above operations also satisfies all the criteria stated in Section 3.1. Thus values greater than 1 are out of the set and less than one are in the set. This means that resulting expressions from one set of operations can be directly fed into the next set of operations without making any changes like complex exponentiation etc. Hence we can find

(AUB) ∩ (AUC) by finding (AUB) and (AUC) individually and then pass the result into intersection operation.

There could be a challenge of dealing with very large numbers as the expressions obtained by the above methods are growing exponentially. We can manipulate the last step in the method to make it asymptotically reach a finite value without compromising the quality of any of the results. The final output can be substituted into the expression $(1 - e^{-\ln(2)*x}) \leq 0.5$ instead of x and the result will be bounded in [0, 2] thus removing any possibility of large numbers as a result. Note that this must be the last step as it does not satisfy the properties needed for performing the set operations. We will not use this method in further topics as it is not needed.

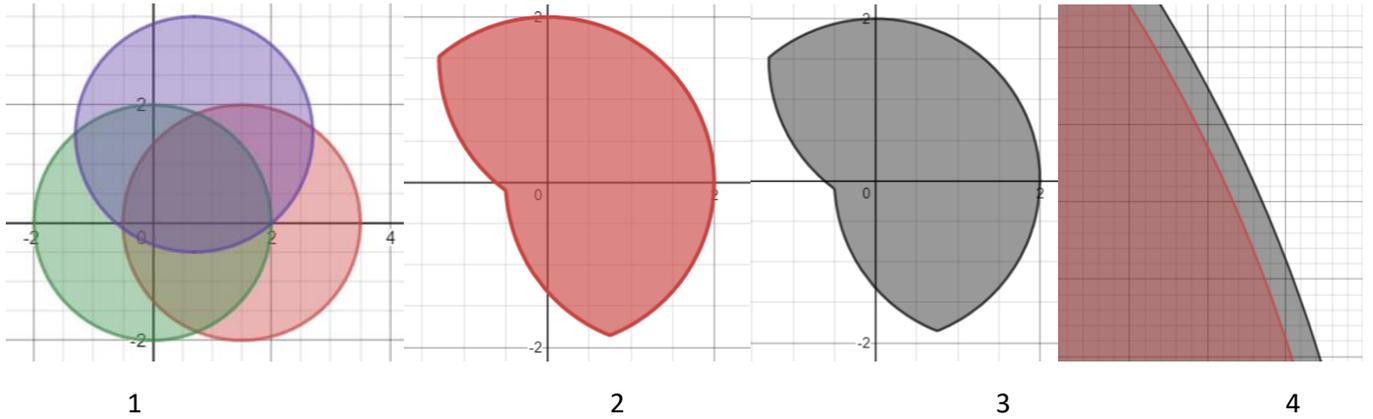

Fig. 9: Distributive Property and its dynamics

*(1) Sets A, B and C from bottom left to top; (2) A ∩ (B∪C) Eq. 12; (3) (A∩B) U (A∩C) Eq. 13; (4) Error when all parameters are 5. Expressions:*

$$A: x^{2} + y^{2} - 4 \leq 0 \quad B: (x - 1.5)^{2} + y^{2} - 4 \leq 0 \quad C: (x - 0.7)^{2} + (y - 1.5)^{2} - 4 \leq 0$$

$$A \cap (B \cup C): e^{\{5(x^{2}+y^{2}-4)\}} + \left(e^{\{-5((x-1.5)^{2}+y^{2}-4)\}} + e^{\{-5((y-1.5)^{2}+(x-0.7)^{2}-4)\}}\right)^{\{-1\}} \leq 1 \quad ...Eq.12$$

$$(A \cap B) U (A \cap C):$$

$$\left(\left(e^{\{5(x^{2}+y^{2}-4)\}} + e^{\{5((x-1.5)^{2}+y^{2}-4)\}}\right)^{\{-1\}} + \left(e^{\{5(x^{2}+y^{2}-4)\}} + e^{\{5((y-1.5)^{2}+(x-0.7)^{2}-4)\}}\right)^{\{-1\}}\right)^{\{-1\}} \leq 1 \quad ...Eq.13$$

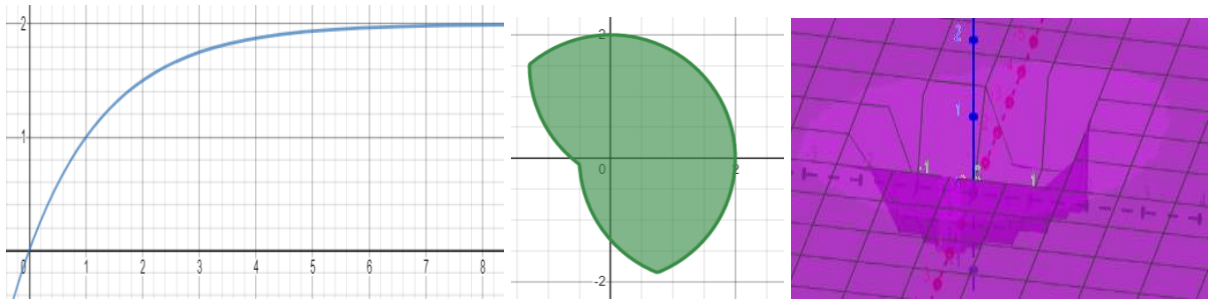

Fig. 10: left to right: $2(1 - e^{-\ln(2)*x}) = y$; Graph of Eq. 14; 3D representation of LHS of Eq. 14:

Bounded Equivalent Expression for $A \cap (B \cup C)$:

$$\left(1 - e^{\left\{-\ln(2)\left(e^{\{50(x^{2}+y^{2}-4)\}} + \left(e^{\{-50((x-1.5)^{2}+y^{2}-4)\}} + e^{\{-50((y-1.5)^{2}+(x-0.7)^{2}-4)\}}\right)^{\{-1\}}\right)\right\}}\right) \leq 0.5 \quad ...Eq.14$$

## 5. Visualisations and Intuitions

The above operations can be visualized and an intuitive understanding for the same can be developed as follows:

Example 1: Functions of the form $f(x,y) \leq 0$ can be shown as:

| A | B | C | D | E | F | G | H |
|---|---|---|---|---|---|---|---|
| x-2 | -(x+2) | y-2 | -(y+2) | $(x-2)^2+y^2-1$ | $(x+2)^2+y^2-1$ | $x^2+(y+2)^2-1$ | $x^2+(y-2)^2-1$ |

Set Expression: $((((A \cap B \cap C \cap D) \cup E) \cap F) \cup G) \cap H$ which result into Eq. 15

Equation:

$$e^{\{a*(-(x^{\{2\}}+(y-2)^{\{2\}}-1))\}} + \left(\left(e^{\{a*(x^{\{2\}}+(y+2)^{\{2\}}-1)\}}\right)^{\{-1\}} + \left(e^{\{a*(-((x+2)^{\{2\}}+y^{\{2\}}-1))\}} + \left(\left(e^{\{a*((x-2)^{\{2\}}+y^{\{2\}}-1)\}}\right)^{\{-1\}} + \right.\right.\right.$$

$$\left.\left.\left.\left(e^{\{a*(-(y+2))\}} + e^{\{a*(y-2)\}} + e^{\{a*(-(x+2))\}} + e^{\{a*(x-2)\}}\right)^{\{-1\}}\right)^{\{-1\}}\right)^{\{-1\}}\right)^{\{-1\}} \leq 1 \quad \ldots Eq.\,15$$

Example 2: Differentiable Batman Equation in Eq. 1:

| A | B | C | D | E |
|---|---|---|---|---|
| -(y+1.5) | (3(y-0.1)-258.18((1.9x+0.1)(1.9x-0.1))^1.6) | -((3(x-0.45))^14-y) | -((3(x+0.45))^14-y) | (y-0.4) |
| **F** | **G** | **H** | **I** | **J** |
| -((0.5(y+1.6))^8+(-x+2)) | -(((0.5(y+1.6))^8+(x+2))) | -((0.5(y+1.6))^8+(x+2)) | (0.5(x+1.16)^2.8)^2+(y+0.6) | (0.5(x-1.16)^2.8)^2+(y+0.6) |

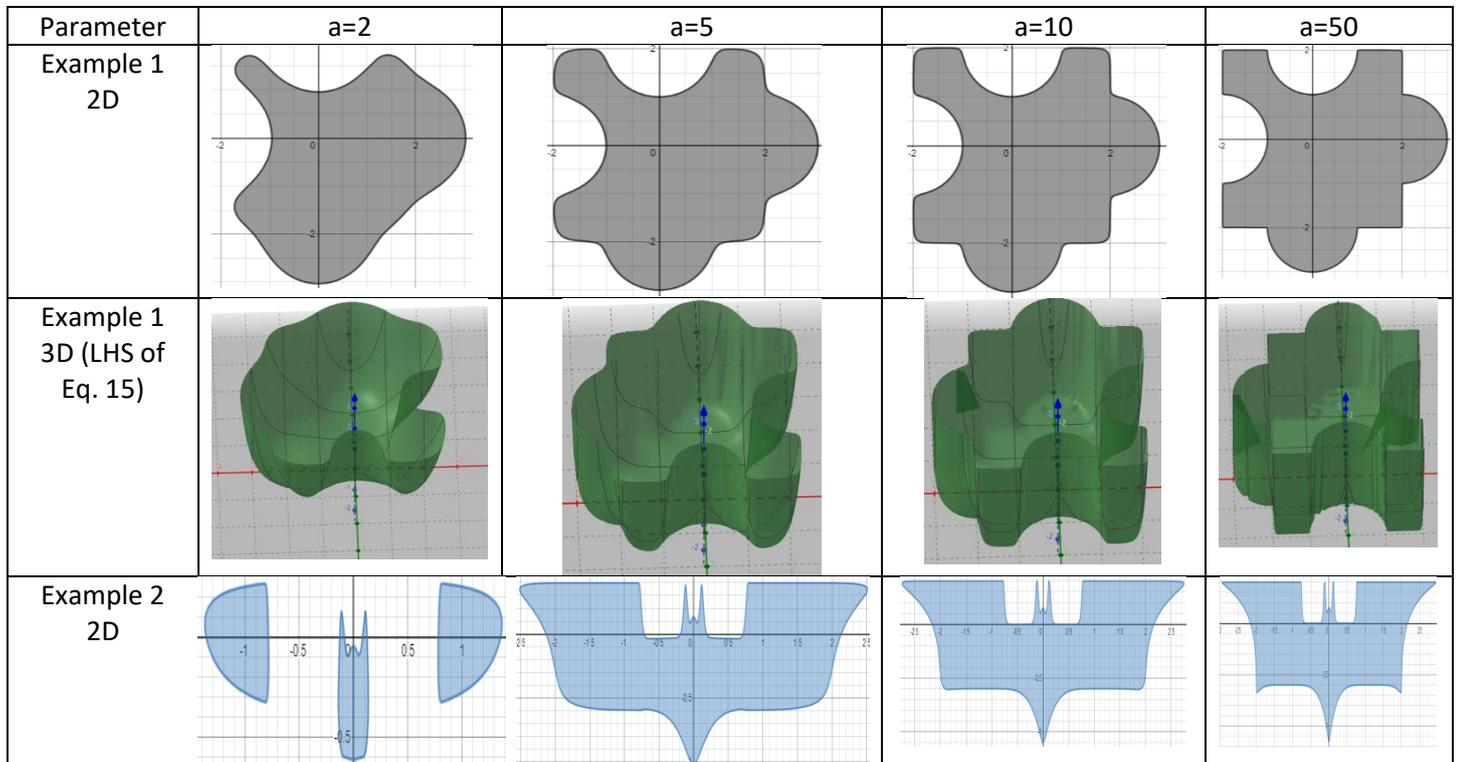

| Parameter | a=2 | a=5 | a=10 | a=50 |
|---|---|---|---|---|
| Example 1 2D | | | | |
| Example 1 3D (LHS of Eq. 15) | | | | |
| Example 2 2D | | | | |

*Visualising the 2 and 3 dimensional space generated by the proposed method. The sharp increase in the values of the expressions is clearly seen in 3d plot. The increase of sharpness of the two examples is also an important observation.*

# 6. Applications

1. **Performing Spline and Rounding of Corners:**

   As shown in the visualisation section, polynomials can be used to implement the spline functionality and their results can be merged using this method. Rounding of corners and edges can also be performed by adjusting the parameter used. Different values of the parameter can be used for different effects.

2. **Deriving Neural Network Activation Function: Softplus from ReLU:**

   An attempt to apply the algorithm on the ReLU[9] activation function used in neural networks was made. It was observed that suggested approach was able to derive another popular activation function known as softplus [10] by applying the same technique as stated above on ReLU.

   Derivation:
   $$ReLU: f(x,y) = y; g(x,y) = y - x$$

   Performing Union we get:
   $$\frac{1}{e^{-ay} + e^{-a(y-x)}} \leq 1$$

   The region simplifies to:
   $$y \leq \frac{\ln(1 + e^{ax})}{a}$$

   On identifying the boundary and substituting a as 1 we obtain the softplus activation function. Note that a general family of softplus activation functions is obtained in which parameter 'a' will control its similarity to ReLU.

| A=1 (Softplus) | a=2 | a=5 | a=10 |
|---|---|---|---|
| 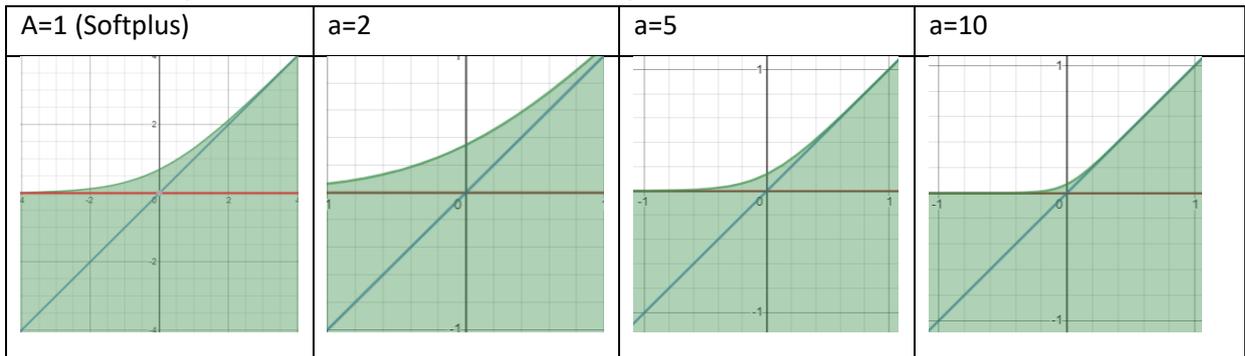 | | | |

*The effect of parameter 'a' on the softplus activation function*

3. **Differentiable equivalent functions for operations like Max() and Min():**

   Computing $\max(\sin(x), x - 5, -x - 5)$ will involve converting all the terms into equations of the form $f(x,y) \geq 0$ followed by performing intersection. Similarly $\min(\sin(x), x + 5, -x + 5, -(x/3)^2 + 10)$ will need intersection without any negations. The following images demonstrate the same with a=10:

   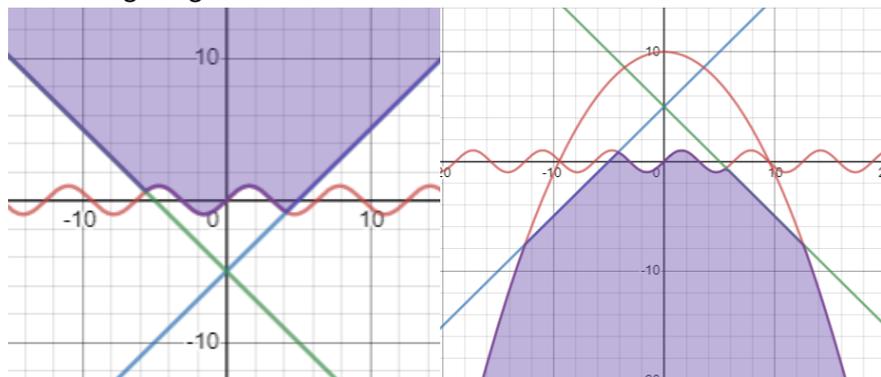

   $$Min(right): e^{\{a(y-\sin(x))\}} + e^{\{a(y-x-5)\}} + e^{\{a(y+x-5)\}} + e^{\left\{a\left(y+\left(\left(\frac{x}{3}\right)\right)^{\{2\}}-10\right)\right\}} \leq 1$$
   $$Max(left): e^{\{-a(y-x+5)\}} + e^{\{-a(y+x+5)\}} + e^{\{-a(y-\sin(x))\}} \leq 1$$

4. Animations and Transitions:

   Animations and transitions can be well simulated for aesthetic purposes in presentation and animation softwares. New appearance and other animations can be created by changing the parameter values and/ or including a dynamic boundary in the equations as shown in the equation and animation instances given below:

$$\left(\left(e^{\{a*(-y)\}} + e^{\{a*(y-5)\}} + e^{\{a*(x-2)\}} + e^{\{a*(1.6-x)\}}\right)^{\{-1\}} + \left(e^{\{-a*(x-2)\}} + e^{\{a*(-y)\}} + e^{\{a*(-(y+x-3.4))\}} + e^{\{a*(y+x-3.8)\}}\right)^{\{-1\}}\right.$$
$$\left. + \left(e^{\{a*-(x-2)\}} + e^{\{a*((x-2)^2+(y-3.3)^2-3)\}} + e^{\{a*-((x-2)^2+(y-3.3)^2-2)\}}\right)^{\{-1\}}\right)^{\{-1\}} \leq 1$$

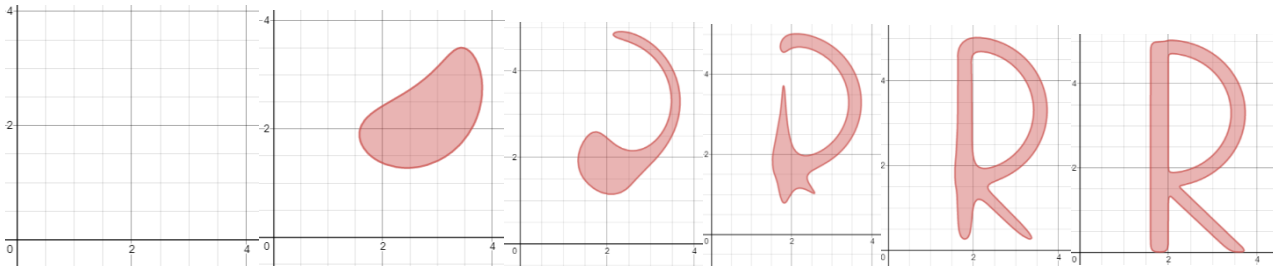

*How the animation would evolve with increase in the value of parameter 'a'*

5. Optical Character Recognition and Pattern Matching:

   We expect uses of this approach in the field of optical character recognition. As the structural features of various characters can be mapped as shown by the inequality $F(x,y) \leq 1$ which can be derived in a way similar to the one demonstrated above, we can use the actual value of the function shown below as the loss function:

$$\max\left(\sum_i^N F(x_i, y_i) - N, 0\right) \text{ where } N \text{ is number of points represented by } (x_i, y_i)$$

   Thus this approach will be able to compute a membership function of the given pattern. Note that some values may have to be clipped as exponentials may grow very fast and the suggested bounding method can also be applied here.

## 7. Future Work

This mathematically very simple and basic approach can have applications ranging from computer graphics, to optimization techniques and also modern machine learning methods. We would like to explore the effects of different parameters if assigned separately to every component. Simultaneously we will continuously explore the applications of this technique in Multi-Linear optimization and development of new optimization techniques and loss functions for Machine and Deep Learning applications.

## 8. Acknowledgements

I would like to acknowledge the work of Bro. Harshmeet Singh Wadhwa from Bishops School Pune for supporting me in perfecting the equations used as the components of the differentiable batman equation. Geogebra and Desmos tools were of utmost importance in the development of this project and hence special credits go to their developers.

## Appendix:

## Python Script to Perform the above Operations:

The resultant of the following script can be directly pasted into Desmos for plotting:

```python
# note that this script parses expressions for the Desmos graphical calculator
# Input set expression must be in postfix form
stk = []
expr = ''
varss = {}
alpha = 'abcdefghijklmnopqrstuvwxyz' #alphabet for parsing
e = ''
sf = input('Enter the sharpness factor: ') # factor a considered same for all terms currently
print('Enter the expressions: ')
while e!='':
    e = input()
    if e!='':
        a,b = e.split(' ') # splitting the set variable and inequality
        varss[a] = 'e^{'+sf+'*('+b+')}' # representing input

def intersection(a,b): # performing intersection
    return a+'+'+b

def union(a,b): # performing union
    a = '('+a+')^{-1}'
    b = '(' + b + ')^{-1}'
    return '('+intersection(a,b)+')^{-1}'

def postfix_parse(expr):
# Postfix parser
# Inverse is not included in the script
    global stk

    for i in range(len(expr)):
        print(i,stk)
        if expr[i] in alpha:
            stk.append(expr[i])
        elif expr[i]=='&':
            if stk[-1] in varss.keys():
                x = varss[stk.pop(-1)]
            else:
                x = stk.pop(-1)
            if stk[-1] in varss.keys():
                y = varss[stk.pop(-1)]
            else:
                y = stk.pop(-1)
            stk.append(intersection(x,y))
        elif expr[i] == '|':
            if stk[-1] in varss.keys():
                x = varss[stk.pop(-1)]
            else:
                x = stk.pop(-1)
            if stk[-1] in varss.keys():
                y = varss[stk.pop(-1)]
            else:
                y = stk.pop(-1)
            stk.append(union(x,y))
        else:
            print('Error at ',i,' Invalid symbol ',expr[i])

    return stk[0]+'\le1
```

## Example:

```
Enter the sharpness factor: 50
Enter the expressions:
a \left(x-2\right)^2+\left(y-3.3\right)^2
b \left(x-2\right)^2+\left(y-3.3\right)^2
c x-2

Enter the expression: abc&&
0 []
1 ['a']
2 ['a', 'b']
3 ['a', 'b', 'c']
4 ['a', 'e^{50*(x-2)}+e^{50*(\left(x-2\right)^2+\left(y-3.3\right)^2)}']
e^{50*(x-2)}+e^{50*(\left(x-2\right)^2+\left(y-3.3\right)^2)}+e^{50*(\left(x-2\right)^2+\left(y-3.3\right)^2)}\le1

Process finished with exit code 0
```